\documentclass[conference]{IEEEtran}
\IEEEoverridecommandlockouts
\usepackage{subcaption}
\usepackage{cite}
\usepackage{amsmath,amssymb,amsfonts}
\usepackage{algorithm}
\usepackage{framed}
\usepackage{bbm}
\usepackage{braket}
\usepackage{algpseudocode}
\usepackage{graphicx}
\usepackage{comment}
\usepackage{textcomp}
\usepackage{xcolor}
\usepackage{orcidlink}
\def\BibTeX{{\rm B\kern-.05em{\sc i\kern-.025em b}\kern-.08em
    T\kern-.1667em\lower.7ex\hbox{E}\kern-.125emX}}
\begin{document}

\title{Crossing The Gap Using Variational Quantum Eigensolver: A Comparative Study}

\author{\IEEEauthorblockN{I-Chi Chen\orcidlink{0009-0008-1622-868X}\textsuperscript{1,2}, Nouhaila Innan\orcidlink{0000-0002-1014-3457}\textsuperscript{3}, Suman Kumar Roy\textsuperscript{4}, and Jason Saroni\orcidlink{0009-0003-7197-2309}\textsuperscript{1,2,5}}
\IEEEauthorblockA{\textsuperscript{1}Iowa State University, Ames, USA\\
\textsuperscript{2}Ames National Laboratory, Ames, Iowa 50011, USA\\
\textsuperscript{3}Quantum Physics and Magnetism Team, LPMC, Faculty of Sciences Ben M'sick,\\ Hassan II University of Casablanca,
Morocco\\
\textsuperscript{4}Information Technology, NITK Surathkal, India\\
\textsuperscript{5}Superconducting Quantum Materials and Systems Center (SQMS),\\ Fermi National Accelerator Laboratory, Batavia, Illinois 60510, USA\\
ichen@iastate.edu, nouhaila.innan-etu@etu.univh2c.ma, roysuman.212it031@nitk.edu.in, jsaroni@iastate.edu}}

\maketitle

\begin{abstract}
Within the evolving domain of quantum computational chemistry, the Variational Quantum Eigensolver (VQE) has been developed to explore not only the ground state but also the excited states of molecules. In this study, we compare the performance of Variational Quantum Deflation (VQD) and Subspace-Search Variational Quantum Eigensolver (SSVQE) methods in determining the low-lying excited states of $LiH$. Our investigation reveals that while VQD exhibits a slight advantage in accuracy, SSVQE stands out for its efficiency, allowing the determination of all low-lying excited states through a single parameter optimization procedure. We further evaluate the effectiveness of optimizers, including Gradient Descent (GD), Quantum Natural Gradient (QNG), and Adam optimizer, in obtaining $LiH$'s first excited state, with the Adam optimizer demonstrating superior efficiency in requiring the fewest iterations. Moreover, we propose a novel approach combining Folded Spectrum VQE (FS-VQE) with either VQD or SSVQE, enabling the exploration of highly excited states. We test the new approaches for finding all three $H_4$'s excited states. Folded Spectrum SSVQE (FS-SSVQE) can find all three highly excited states near $-1.0$ Ha with only one optimizing procedure, but the procedure converges slowly. In contrast, although Folded spectrum VQD (FS-VQD) gets highly excited states with individual optimizing procedures, the optimizing procedure converges faster.
 \end{abstract}


\section{Introduction}
\begin{figure}[htpb]
    \centering
    \includegraphics[width=1\linewidth]{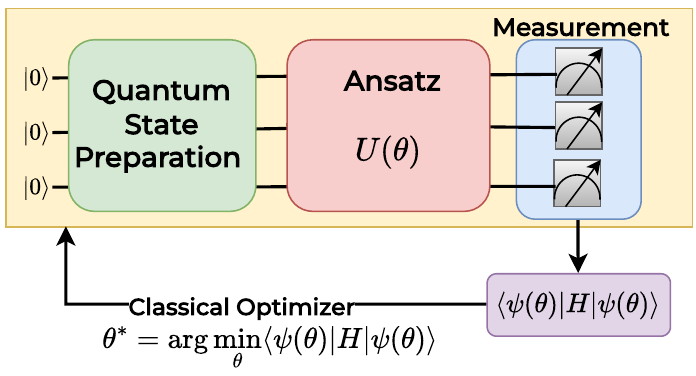}
    \caption{Schematic representation of the VQE process. Qubits are initialized to $|0\rangle$ during Quantum State Preparation. A parameterized quantum circuit, denoted as ansatz  $U(\boldsymbol{\theta})$, is applied to the qubits to prepare the trial state $|\psi(\boldsymbol{\theta})\rangle$. The system undergoes measurement to retrieve information used to evaluate the expectation value of the Hamiltonian $H$. A classical optimizer processes the measurement results to find the optimal parameters $\boldsymbol{\theta}^*$, which minimize the cost function, corresponding to the ground state energy of the Hamiltonian.}
    \label{vqearch}
\end{figure}

With the rapid development of Quantum Computing (QC), there are more and more applications using QC in chemistry\cite{Cao2019, Higgott2019, Nakanishi2019, McArdle2020, Chan2021, Tazi2024, INNAN2024}, biology\cite{Marx2021, Cordier2022}, high energy physics\cite{Meth2023}, and quantum many-body dynamics physics\cite{Smith2019, Mi2022, Chen2022, Shtanko2023, Lamm2023, Chen2023}. However, compared to a decade ago, QC hardware is much improved, but the error is still too large for quantum error correction. The quantum hardware with lots of noise (gates' error rate around $1 \times 10^{-2}$) and a limited number of qubits (around $10$ to $ 1000$ physical qubits) is called Noisy Intermediate Scale Quantum (NISQ) devices \cite{Preskill18}. While NISQ devices cannot perform complex error correction yet, they promise to solve specific problems in areas like finance\cite{Pistoia2021,Herman2022} and public transportation\cite{Bentley2022}. 
One of the cutting-edge algorithms, which are friendly for this NISQ device, is Variational Quantum Eigensolver (VQE)\cite{Peruzzo2014,kandala2017hardware,tilly2022variational,cerezo2022variational}. VQE is a hybrid quantum algorithm that gets the lowest eigenstate with a given Hamiltonian ($H$) using variational ansatz. As fig.~\ref{vqearch} shows, initially, the trial state $\ket{\psi(\theta)}=U\left(\theta\right)\ket{\psi}$ with initial state $\ket{\psi}$ is generated by a parameterized quantum circuit represented by $U(\theta)$.  After the measurement for getting the Hamiltonian's expectation value, the classical optimizer optimizes the loss function, which is also the expectation value of the Hamiltonian
\begin{equation}    
\mathcal{L}(\boldsymbol{\theta})=\bra{\psi(\boldsymbol{\theta})}H\ket{\psi(\boldsymbol{\theta})},
\label{eq: H}
\end{equation}
by adjusting the parameters. The state with optimal parameters $\boldsymbol{\theta}^*$, which has minimum $\braket{H}$, is approximately the ground state. With its streamlined circuit design, VQE shows great promise for advancing quantum computation in the foreseeable future.

However, VQE with the loss function in eq.~\ref{eq: H} is only designed for solving the lowest state and its energy. There are many VQEs with specific loss functions to get the excited states. The paper \cite{Higgott2019} adds the fidelity between target excited states and lower energy states to the loss function so that the target state will be the excited state with energy higher than other lower energy states. The corresponding VQE technique is called Variational Quantum Deflation (VQD).
Moreover, the research \cite{Nakanishi2019} takes the sum of different orthogonal states' Hamiltonian expectation values as the loss function. 
Thus, obtaining both the ground and lower excited states within a single parameter optimization procedure is possible. The approach is known as Subspace-Search Variational Quantum Eigensolver (SSVQE). In this work, we focus on these two methods and test their efficacy in gaining the ground state and excited states of molecules. Nevertheless, these two approaches are only suitable for finding low-lying excited states. Hence, we propose a method that combines VQD or SSVQE with the fold spectrum technique \cite{Tazi2024}. This method enables the exploration of the spectrum and facilitates the identification of excited states near specific energy levels.

This paper is organized as follows. In Sec.~\ref{sec2}, we present our methodology, including mapping quantum problems to Pauli operators, an overview of various VQE approaches, the design of ansatzes, qubit tapering techniques, and the choice of classical optimizers. 
Sec.~\ref{sec3} discusses the results of SSVQE and VQE, the comparison among optimizers, and the result of combining VQE techniques, which involve FS-VQE and VQD or SS-VQE, for calculating highly exciting states near the specific energy level. 
Finally, Sec.~\ref{sec4} provides a concise conclusion, summarizing the key takeaways and suggesting directions for future research.

\section{Methodology\label{sec2}}
\subsection{Mapping into Pauli Operator}
The general chemical molecule's Hamiltonian in terms of the second quantization is given by
\begin{equation}
H=\sum_{i,j}h_{i,j}a_{i}^{\dagger}a_{j}+\frac{1}{2}\sum_{ijkl}h_{ijkl}a_{i}^{\dagger}a_{j}^{\dagger}a_{k}a_{l},
\label{eq: Ham}
\end{equation}
where $a^{\dagger}_i$, ($a_i$) is the creation (annihilation) operator for the spin-orbital $i$, and $h_{i,j}$ ($h_{i,j,k,l}$) are the one (two) electrons integral. 

In order to simulate the Hamiltonian on quantum computers,  it is crucial to map electronic operators into Pauli operators. The general way is Jordan Wigner's mapping
\begin{equation}
\begin{aligned}
a_{j}^{\dagger} &= \frac{1}{2}\left(X_{j}-iY_{j}\right)\prod_{k<j}Z_{k}, \\
a_{j} &= \frac{1}{2}\left(X_{j}+iY_{j}\right)\prod_{k<j}Z_{k},
\end{aligned}
\label{eq: JW}
\end{equation}

where $Z_i$, $X_i$, and $Y_i$ are Pauli operators on spin-orbital $i$. After the mapping, the Hamiltonian in eq.\ref{eq: Ham} becomes
\begin{equation}
H=\sum_{l}\eta_{l}P_{l},
\label{eq: Ham2}
\end{equation}
where $P_{l}\in\left\{ I,X,Y,Z\right\} ^{\bigotimes M}$ are the Pauli strings with $M$ the total number of qubits, and $\eta_{l}$ are the corresponding component of Pauli string. The Hamiltonian in the form of eq.~\ref{eq: Ham2} can generated by the PennyLane module \textbf{qchem.molecular\_hamiltonian}.

\subsection{Different Types of VQE}
\subsubsection{Subspace-Search Variational Quantum Eigensolver}
SSVQE is an algorithm designed to tackle the challenge of calculating excited states \cite{Nakanishi2019}. It efficiently explores a low-energy subspace to search for the k-th excited state by utilizing orthogonal input states $\left\{ \ket{\psi_{i}}\right\} $ and applying common variational unitary transformations to the subspace, which is composed of $k$ lowest eigenstates. Original SSVQE requires two unitary transformations to search $k$-th excited state. One, $U(\boldsymbol{\theta})$, is to map the $k$ input states $\left\{ \ket{\psi_{i}}\right\} $ to the superposition of $k$ lowest excited states by minimizing the loss function
\begin{align}
\mathcal{L}_{1}(\boldsymbol{\theta})= \sum_{j=0}^{k}\bra{\psi_{j}}U^{\dagger}(\boldsymbol{\theta})H(\boldsymbol{\theta})\ket{\psi_{j}}.
\label{eq: ssvqel1}
\end{align}
Another $V(\boldsymbol{\phi})$, which only acts on $\left\{ \ket{\psi_{i}}\right\}$, is to transform one of $\left\{ \ket{\psi_{i}}\right\}$ to the superposition of $\left\{ \ket{\psi_{i}}\right\}$ so that $U(\boldsymbol{\theta})$ can transform the superposition state to $k$-th excited state. To get the $k$-th excited state, we can maximize an alternative loss function 
\begin{align}
\mathcal{L}_{2}(\boldsymbol{\phi})=\sum_{j=0}^{k}\bra{\psi_{j}}V^{\dagger}(\boldsymbol{\phi})U^{\dagger}(\boldsymbol{\theta}^{*})HU(\boldsymbol{\theta}^{*})V(\boldsymbol{\phi})\ket{\psi_{j}},
\end{align}
where $\boldsymbol{\theta}^{*}$ means the optimal parameters in eq.~\ref{eq: ssvqel1}. However, it's hard to find a specific ansatz $V(\boldsymbol{\phi})$ that acts only on $\left\{ \ket{\psi_{i}}\right\}$ and requires two optimization processes. The paper~\cite{Nakanishi2019} also proposed weighted SSVQE, which can acquire all excited states up to the k-th excited state through a single parameter optimization procedure. The corresponding loss function is designed as 
\begin{align}
\mathcal{L}_{\boldsymbol{w}}(\boldsymbol{\theta}) = \sum_{j=0}^{k}w_j\bra{\psi_j}U^{\dagger}(\boldsymbol{\theta})HU(\boldsymbol{\theta})\ket{\psi_j},
\end{align}
where the value of weight $w_i \in (1,0)$ is chosen to be smaller and smaller as $i$ index increases.

\subsubsection{Variational Quantum Deflation}

\begin{figure}[t]
     \includegraphics[width=0.45\textwidth]{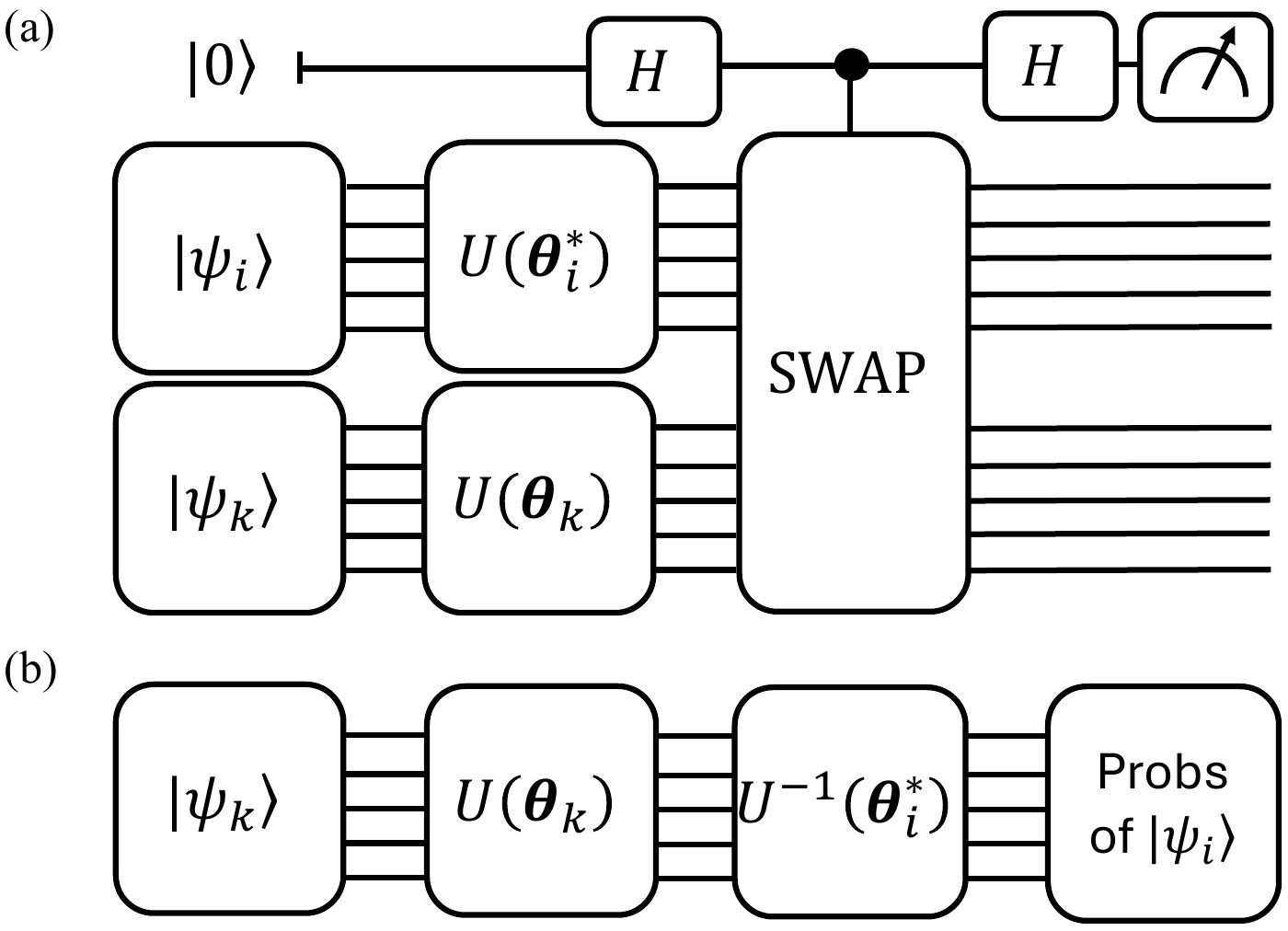}
     \caption{The circuits for calculating the overlap $|\braket{\psi_k(\boldsymbol{\theta}_k)|\psi_i(\boldsymbol{\theta}^*_i)}|^2$. \textbf{(a)} shows the circuit using a swap test; \textbf{(b)} shows the circuit using inversion of unitary to calculate the overlap, with ``Probs'' denoting the probability of getting state $\ket{\psi_i}$, we can measure the circuits in the orthogonal basis that contains $\ket{\psi_i}$.}
     \label{fig:fid}
 \end{figure}

Unlike SSVQE, which has a common parameterized unitary for all orthogonal input states, VQD has different parameterized unitaries $U_k(\boldsymbol{\theta}_k)$ for all input initial states $\left\{ \ket{\psi_{k}}\right\} $, which are not orthogonal with each other and can be all identical. The key to getting low-lying excited states is adding overlap between the training state and other low-lying excited states to the loss function
\begin{align}
\label{eq: VQDL}
\mathcal{L}(\boldsymbol{\theta}_k) = & \bra{\psi_k(\boldsymbol{\theta}_k)}H\ket{\psi_k(\boldsymbol{\theta}_k)}+\sum_{i=0}^{k-1} \beta_i |\braket{\psi_k(\boldsymbol{\theta}_k)|\psi_i(\boldsymbol{\theta}^*_i)}|^2,
\end{align}
where $\ket{\psi_k(\boldsymbol{\theta}_k)}=U_k(\boldsymbol{\theta}_k)\ket{\psi_k}$, $\boldsymbol{\theta}^*_i$ denotes the optimal parameters for getting $i$ th lowest energy state, and $\beta_i$ is chosen to be larger than the energy discrepancy between $i$ th and $i-1$ th lowest energy states. The last term, the overlap term known as the fidelity, can be realized in the quantum circuit shown in fig.~\ref{fig:fid}. As fig.~\ref{fig:fid} shown, the swap test circuit can be utilized, requiring additional qubits to prepare another state and one ancilla qubit to obtain the overlap value. Alternatively, the circuit depicted in fig.\ref{fig:fid} (b) can be employed, utilizing the inversion of the unitary to calculate the overlap. The trick to acquire $k$ lowest energy states using VQD is to optimize the loss function in eq.~\ref{eq: VQDL} individually from the lowest state to $k$ th lowest state.

\subsubsection{Folded Spectrum VQE}
The FS technique is a method to map the Hamiltonian's target highly excited states to the ground state by changing the Hamiltonian
\begin{align}
H'=\left(H-\omega\right)^{2},
\label{eq: fold}
\end{align}
where $\omega$ is an arbitrary scaler energy value close to the excited state's energy level. FS-VQE is a normal VQE to the ground state of $H'$. While FS-VQE has been acknowledged, its quantum application was previously considered too costly due to the exponential growth of terms in the measured operator. Nevertheless, the study's \cite{Tazi2024} implementation reveals a significant advancement by employing a Pauli grouping technique, which can significantly reduce the number of required measurements and makes FS-VQE a cost-efficient option, especially for the second quantized molecular Hamiltonians.

\subsection{The Ansatz}

\subsubsection{Unitary Coupled-Cluster Singles and Doubles Ansatz}
The Unitary Coupled-Cluster Singles and Doubles (UCCSD) ansatz \cite{Lee_2018} is a simplified version of Unitary Coupled-Cluster (UCC) ansatz, which includes all possible fermionic excitations and is described by $U(\boldsymbol{\theta})=e^{T(\boldsymbol{\theta})-T^\dagger(\boldsymbol{\theta})}$ with excitation operators 
\begin{align}
T(\boldsymbol{\theta})-T^\dagger(\boldsymbol{\theta}) &= \sum_{j,a}\theta_j^a(a_j^\dagger a_a-a_a^\dagger a_j) \\ \nonumber
&+\sum_{i,j,a,b}\theta_{ij}^{ab}(a_i^\dagger a_j^\dagger a_a a_b-a_a^\dagger a_b^\dagger a_i a_j) \\ \nonumber
&+\cdots,
\end{align}
where ${(i, j)}$ represents occupied orbitals, ${(a, b)}$ denotes unoccupied orbitals, and ``$\cdots$'' means the higher order excitations. Instead of involving all excitations, UCCSD only includes fermionic single excitation and double excitation. The corresponding ansatz are $U(\boldsymbol{\theta})=e^{T'(\boldsymbol{\theta})-T'^\dagger(\boldsymbol{\theta})}$ with skew-Hermitian operator
\begin{align}
T'(\boldsymbol{\theta})-T'^\dagger(\boldsymbol{\theta}) &= \sum_{r,p}\theta_r^p(a_p^\dagger a_r-a_r^\dagger a_p) \\ \nonumber
&+\sum_{p,q,r,s}\theta_{pq}^{rs}(a_p^\dagger a_q^\dagger a_r a_s-a_r^\dagger a_s^\dagger a_p a_q),
\label{eq: UCCSD}
\end{align}
where the indexes $\{p, q, r, s\}$ can refer to any orbital within the molecule, regardless of whether it is occupied. 

However, the unitary operator $U(\boldsymbol{\theta})$ contains a single exponent, and cannot be directly implemented on a quantum computer. Instead, the Trotter formula is employed to approximate this unitary operation $e^{A+B}\approx(e^{\frac{A}{\Delta}}e^{\frac{B}{\Delta}})^\Delta$. With the first order trotterization ($\Delta = 1$), UCCSD ansatz becomes
\begin{align}
U(\boldsymbol{\theta})\approx\prod_{r,p}\exp\left[\theta_{p}^{r}\tau_{pr}^{(s)}\right]\prod_{p,q,r,s}\exp\left[\theta_{pq}^{rs}\tau_{pqrs}^{(d)}\right],
\end{align}
where $\tau_{pr}^{(s)}$, $\tau_{pqrs}^{(d)}$ are the fermionic single and double excitation operators
\begin{align}
\tau_{pr}^{(s)}=a_{p}^{\dagger}a_{r}-a_{r}^{\dagger}a_{p},
\end{align}
\begin{align}
\tau_{pqrs}^{(d)}=a_{p}^{\dagger}a_{q}^{\dagger}a_{r}a_{s}-a_{r}^{\dagger}a_{s}^{\dagger}a_{p}a_{q}.
\end{align}
Nonetheless, UCCSD with fermionic single and double excitation operators excitation requires many CNOT gates to implement. This makes it challenging to implement the UCCSD ansatz on NISQ devices.

\subsubsection{Qubit Coupled Cluster Singles and Doubles Ansatz}
In order to reduce gate number for NISQ device, the study~\cite{Yordanov2020} proposed the Qubit Coupled Cluster Single and Double (QCCSD) ansatz. Instead of using fermionic single and double excitation operators, QCCSD utilizes the single and double qubit excitation operators
\begin{align}
\tilde{\tau}_{ik}^{(s)}=Q_{i}^{\dagger}Q_{k}-Q_{k}Q_{i},
\label{eq: qse}
\end{align}
\begin{align}
\tilde{\tau}_{ijkl}^{(d)}=Q_{i}^{\dagger}Q_{j}^{\dagger}Q_{k}Q_{l}-Q_{k}^{\dagger}Q_{l}^{\dagger}Q_{k}Q_{l},
\label{eq: qsd}
\end{align}
where $Q_{i}$ ($Q_{i}^{\dagger}$) is the qubit annihilation (creation) operator and can also be written in terms of Pauli operators
\begin{align}
Q_{j}=\frac{1}{2}\left(X_{j}+iY_{j}\right),
\end{align}
\begin{align}
Q^\dagger_{j}=\frac{1}{2}\left(X_{j}-iY_{j}\right).
\end{align}
With single and double qubit excitation operators, the first order trotterized QCCSD ansatz is given by
\begin{align}
U(\boldsymbol{\theta})=\prod_{i,k}\exp\left[\theta_{ik}\tilde{\tau}_{ik}^{(s)}\right]\prod_{i,j,k,l}\exp\left[\theta_{ijkl}\tilde{\tau}_{ijkl}^{(d)}\right],
\end{align}
where the first (second) term on the right-hand side is the single (double) qubit excitation evolution operator. Unlike single and double fermionic excitation evolution operators of which the number of CNOT gates needed is system size dependent, implementing these single qubit (double) evolution operators on a quantum computer requires 2 (13) CNOT gates for any system size \cite{Yordanov2021}. 

To calculate the excited states more efficiently, spin symmetry and electron number conservation can be used to restrict the trial wavefunction, conserving total spin and electron number. In this work, we also implement the spin restriction on the QCCSD ansatz for our simulation. Thus, the QCCSD ansatz employed in our simulations contains only single and double qubit excitation evolution operators that conserve the total spin number.

\begin{figure}[t]
     \includegraphics[width=0.5\textwidth]{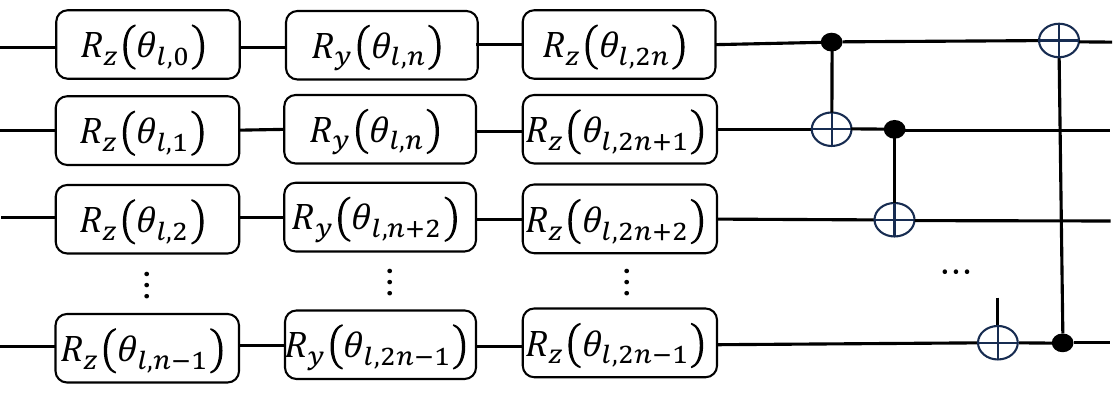}
     \caption{A layer of the quantum circuit for a strongly entangled variational ansatz. It consists of n qubits with alternating layers of parameterized $R_z$ and $R_y$ gates, followed by nearest-neighbor CNOT gates for entanglement. The parameters $\theta_{l,k}$ are optimized for the eigenstate preparation.}
     \label{fig:ansatz}
 \end{figure}
\subsubsection{Strongly Entangled Ansatz}
The Strongly Entangled (SE) ansatz consists of layers of gates, as fig.\ref{fig:ansatz} shows. Each layer is composed of $1$-qubit rotational gates with followed CNOT gates, which make nearest neighbor qubits entangle with each other \cite{Schuld2020}. Although the gates are native for most NISQ devices, the ansatz doesn't conserve the total spin number and charge. It means that the excited state energies obtained using this ansatz are unphysical. To avoid unphysical results, we should add a constraint to the cost function. One of the constraints for conserving the total spin is
 \begin{equation}
    \left\langle \psi\left(\boldsymbol{\theta}\right)\left|\left(S_{z}-m_{z}\right)^{2}\right|\psi\left(\boldsymbol{\theta}\right)\right\rangle,
 \label{eq:sz}
\end{equation}
where $S_z$ is the total magnetization operator, and $m_z$ is the desired result for the total magnetization.

\subsection{Qubit Tapering} 
Given the Hamiltonian in the form of eq.~\ref{eq: Ham2}, the qubit tapering technique allows for the omission of qubits on which, from all the terms $P_{l}$ in the Hamiltonian, at most one Pauli operator acts\cite{Bravyi17,Setia20}. The Pauli operators in the Hamiltonian can be substituted by the Pauli operators' eigenvalue $\pm1$ so the corresponding qubits can be neglected. To reduce the number of qubits, the unitary transform can be used to get the Hamiltonian $H'=UHU^\dagger$ with the largest subset of qubits that is acted trivially or by at most one of the Pauli operators from all the $P_l$ terms.

\subsection{Optimizer}

\subsubsection{Gradient Descent}
Gradient Descent (GD) \cite{ruder2017overview} is an essential optimization algorithm utilized in machine learning and numerical optimization to minimize a loss function by iteratively moving in the direction of the steepest descent of the gradient. For VQE, GD is utilized to optimize the parameters of a quantum circuit, representing the trial wavefunction used to estimate the system's ground state energy. To execute GD, it is necessary to compute the gradient of the loss function $\mathcal{L}(\boldsymbol{\theta})$ concerning the parameters $\boldsymbol{\theta}$. The objective is to minimize the expectation value of the Hamiltonian $H$ with respect to specified quantum state $\ket{\psi(\boldsymbol{\theta})}$ that is defined by the variational parameters $\boldsymbol{\theta}$ and this task involves minimizing the objective function $\mathcal{L}(\boldsymbol{\theta}) = \braket{\psi(\boldsymbol{\theta})|H|\psi(\boldsymbol{\theta})}$. This process entails evaluating the gradient of the expectation value: 

\begin{equation}
\nabla_{\boldsymbol{\theta}} \mathcal{L}(\boldsymbol{\theta})=\braket{\psi(\boldsymbol{\theta})|(H-\mathcal{L}(\boldsymbol{\theta}))\dfrac{\delta U(\boldsymbol{\theta})}{\delta \boldsymbol{\theta}}|\psi(\boldsymbol{\theta})},   
\end{equation}

where $\dfrac{\delta U(\boldsymbol{\theta})}{\delta \boldsymbol{\theta}}$ represents the derivative of the quantum circuit $U(\boldsymbol{\theta})$ with respect to $\boldsymbol{\theta}$. By leveraging the computed gradient $\nabla_{\boldsymbol{\theta}} E(\boldsymbol{\theta})$, GD is implemented to iteratively update the parameters $\boldsymbol{\theta}: \boldsymbol{\theta}_{new} = \boldsymbol{\theta}_{old} - \eta \nabla_{\boldsymbol{\theta}} E(\boldsymbol{\theta}_{old}),$ where $\eta$ denotes the learning rate. The process involves repeating the evaluation of the energy and gradient, followed by the parameter update until the convergence criteria are satisfied (e.g., minimal change in energy or reaching the maximum number of iterations).

\subsubsection{Quantum Natural Gradient}

The landscape of optimization problems encountered in VQE applications is characteristically intricate, often riddled with many local minima. This complexity underscores the necessity of employing an effective optimization strategy pivotal for the algorithm's successful convergence to the ground state energy of the system under study.

Among various optimization techniques, the Quantum Natural Gradient (QNG) optimization strategy stands out by using the geometric properties of the parameter space \cite{QNG,Wierichs2020}. Contrary to the traditional GD method, which operates under the assumption of an Euclidean metric space, QNG employs the Fubini-Study metric tensor, denoted as $g$, to modulate the optimization step sizes. This tensor captures the variational state space's inherent curvature, facilitating more informed and efficacious optimization steps.

The essence of the QNG approach is encapsulated in the update rule:

\begin{equation}
\boldsymbol{\theta}_{\text{new}} = \boldsymbol{\theta}_{\text{old}} - \eta g(\boldsymbol{\theta}_{\text{old}})^{-1} \nabla f(\boldsymbol{\theta}_{\text{old}}),
\end{equation}

where $\boldsymbol{\theta}$ denotes the variational circuit parameters, $\eta$ signifies the learning rate, $\nabla f(\boldsymbol{\theta})$ is the gradient of the objective function with respect to $\boldsymbol{\theta}$, and $g(\boldsymbol{\theta})$ represents the quantum geometric tensor, also known as Fubini-Study metric tensor. When a quantum parameterized circuit consists of $L$ non commuting layers of unitaries, the corresponding variational state is
\begin{equation}
\ket{\psi\left(\boldsymbol{\theta}\right)}=\prod_{l=1}^{L}\left[\prod_{j=1}^{n}e^{-iA_{l,j}\theta_{l,j}}V_{l}\right]\ket{0},
\end{equation}
where $V_l$ is the $l$th layer's non parametric unitary, $A_{l.j}$ are the generators of the gates, $n_l$, $\theta_{l,j}$ denote the total number of parameters and the $j$th parameter respectively in $l$th non-commuting layer of unitary. The corresponding quantum geometric tensor $g(\boldsymbol{\theta})$ is a $N \times N$ block diagonal matrix
\begin{equation}
g\left(\boldsymbol{\theta}\right)=\left(\begin{array}{cccc}
g^{(1)} & 0 & \cdots & 0\\
0 & g^{(2)} & \cdots & 0\\
\vdots & \vdots & \ddots & \vdots\\
0 & 0 & \cdots & g^{(L)}
\end{array}\right),
\end{equation}
where $N=\sum_{l}^{L}n_{l}$ and $g^{(l)}$ is $l$th layer's $n \times n$ submatrix. The submatrix can be evaluated by the quantum device 
\begin{equation}
g_{i,j}^{\left(l\right)}=\left\langle \psi_{l}\left|A_{l,i}A_{l,j}\right|\psi_{l}\right\rangle -\left\langle \psi_{l}\left|A_{l,i}\right|\psi_{l}\right\rangle \left\langle \psi_{l}\left|A_{l,j}\right|\psi_{l}\right\rangle,
\label{eq: GM}
\end{equation}
with $\ket{\psi_{l'}}$ meaning the variational state acted only first $l'-1$ layers non-commuting unitaries
\begin{equation}
\ket{\psi_{l'}}=\prod_{l=1}^{l'-1}\left[\prod_{j=1}^{n}e^{-iA_{l,j}\theta_{l,j}}V_{l}\right]\ket{0}.
\end{equation}

By accounting for the parameter space's geometry, the QNG optimizer significantly enhances the efficiency of the optimization process. It navigates the circuit's sensitivity to parameter variations, circumventing suboptimal pathways often pursued by conventional optimization methods.

 \subsubsection{Adam}
 Adam is an extension of the Stochastic Gradient Descent (SGD) optimization algorithm\cite{Kingmai14}. It combines ideas from momentum-based methods and  Root Mean Square Propagation (RMSprop\cite{Tieleman2012}) to achieve efficient optimization. The optimizer updates the parameters with adaptive learning rates, the first and second moments for $h$ th iteration
\begin{equation}
\boldsymbol{\theta}_{\text{new}}=\boldsymbol{\theta}_{\text{old}}-\eta_{\text{new}}\frac{\boldsymbol{m}_{\text{new}}^{(1)}}{\sqrt{\boldsymbol{m}_{\text{new}}^{(2)}}+\epsilon},
\end{equation}
with learning rate $\eta$, moments $m^{(1)}, m^{(2)}$, and $\epsilon$ to avoid division of zero. The moment's update rule
\begin{equation}
\eta_{\text{new}}=\eta_{\text{Initial}}\frac{\sqrt{\left(1-\beta_{2}^{h}\right)}}{\sqrt{\left(1-\beta_{1}^{h}\right)}},
\end{equation}
\begin{equation}
\boldsymbol{m}_{\text{new}}^{(1)}=\beta_{1}\boldsymbol{m}_{\text{old}}^{(1)}+\left(1-\beta_{1}\right)\left(\nabla f(\boldsymbol{\theta}_{\text{old}})\right),
\end{equation}
\begin{equation}
\boldsymbol{m}_{\text{new}}^{(2)}=\beta_{2}\boldsymbol{m}_{\text{old}}^{(2)}+\left(1-\beta_{2}\right)\left(\nabla f(\boldsymbol{\theta}_{\text{old}})\right)^{\odot2},
\end{equation}
where $f^{\odot2}$ represents element wise square operation. At initial step, the moments $\boldsymbol{m}^{(1)}, \boldsymbol{m}^{(2)}$ is set to be $0$.

 \begin{figure}[t]
     \includegraphics[width=0.45\textwidth]{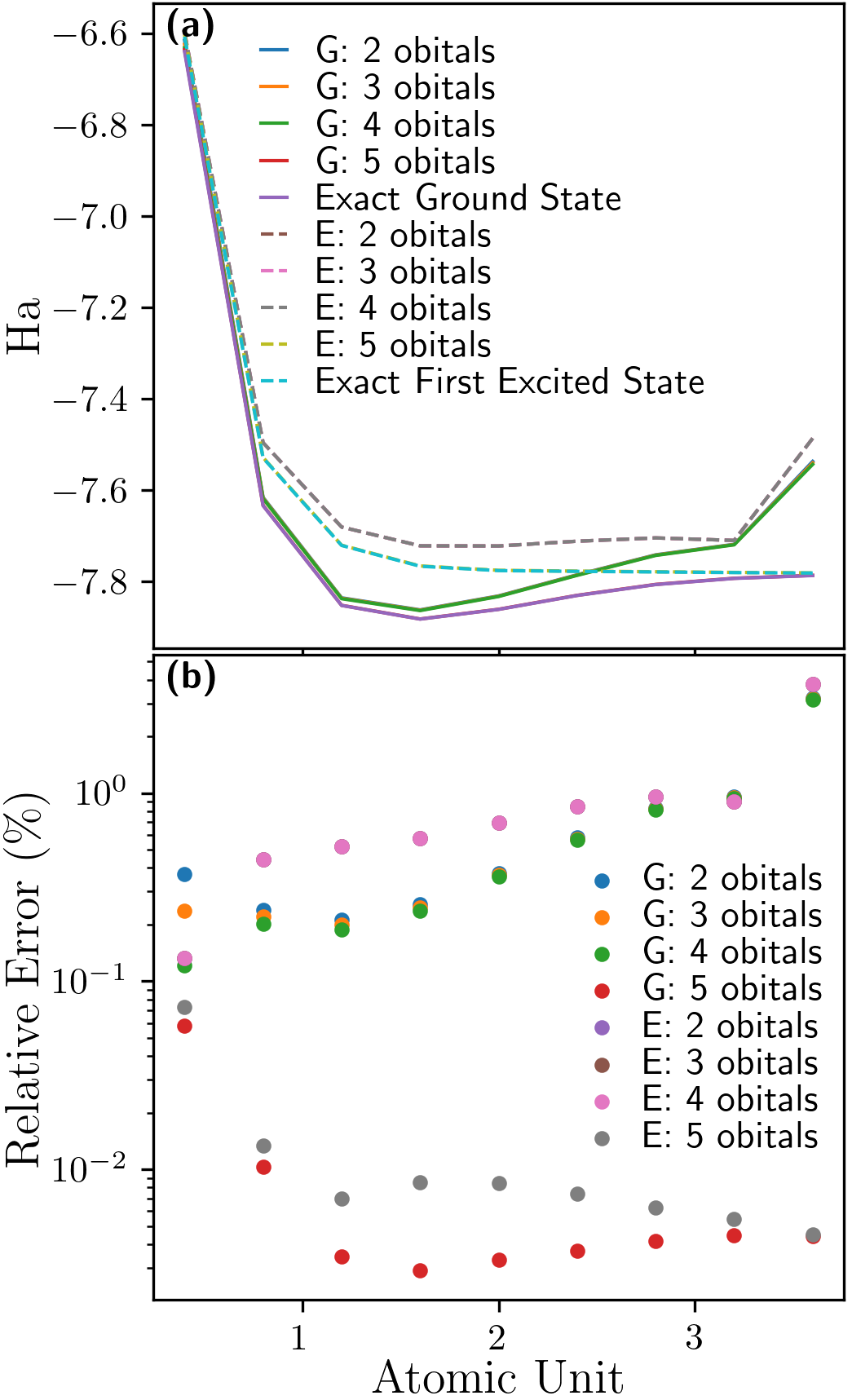}
     \caption{\textbf{(a)} The LiH's energy calculations for the ground state (G) and the first excited state (E) using exact diagonalization with up to 5 active orbitals. Solid and dashed lines represent calculated energies for the ground and excited states, respectively, compared to the exact solutions, and \textbf{(b)} the relative errors of the ground and first excited state energies as a function of atomic units.}
     \label{fig:Orb_Re}
 \end{figure}

\section{Results\label{sec3}}

\subsection{Quantum Resource}
\begin{figure}[t]
     \includegraphics[width=0.5\textwidth]{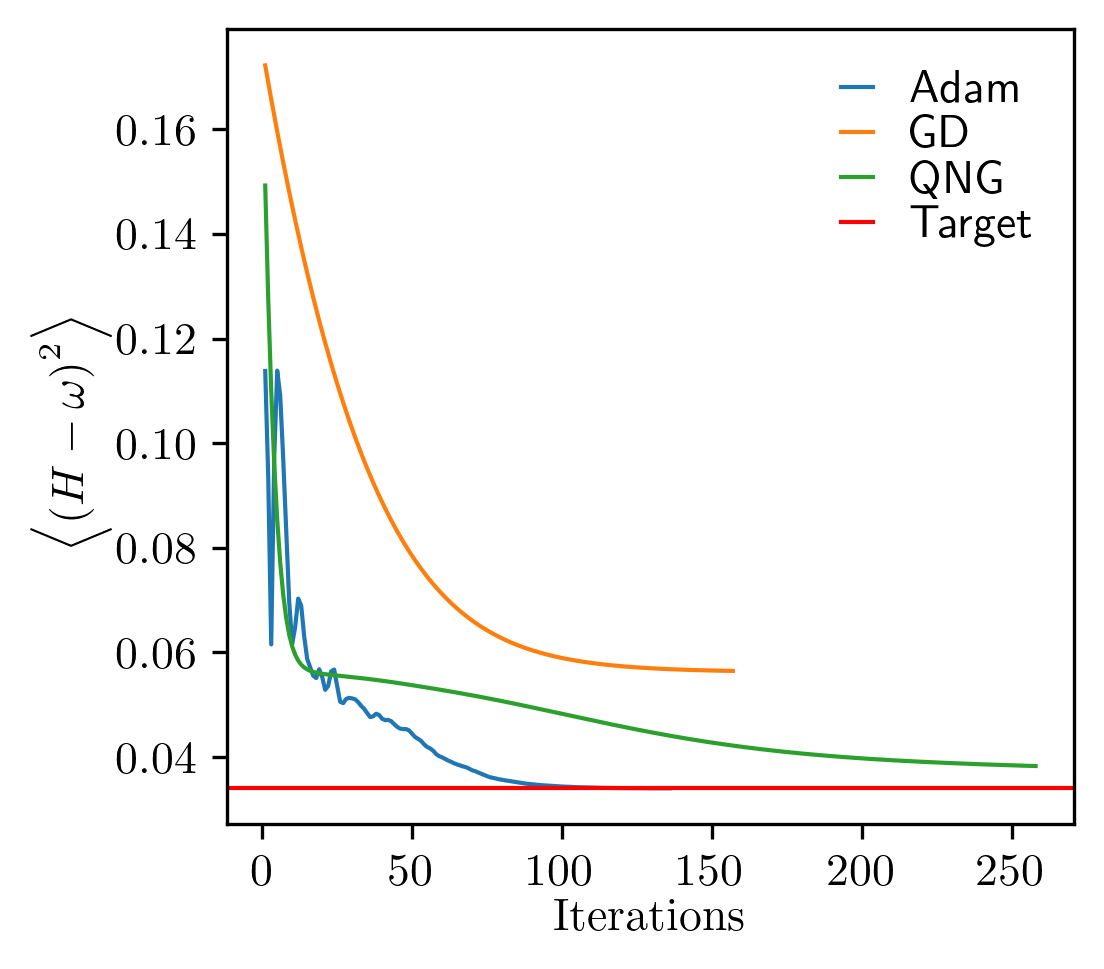}
     \centering
     \caption{The loss function versus the iteration using GD (orange line), QNG (green line), Adam (blue line), with red line as the target value.  }
     \label{fig:OPT}
 \end{figure}


The molecular compound $LiH$ has $6$ molecular orbitals, and each molecular orbital requires 2 qubits to simulate. The number of qubits for the $LiH$'s VQE calculation is $12$. However, we can neglect some inactive molecular orbitals to reduce the number of qubits for VQE simulation. Thus, before VQE simulation, we test how many molecular orbitals we need to get $99.9 \%$ accuracy for the ground state and the first excited state's energy using exactly diagonalizing the Hamiltonian in eq.~\ref{eq: Ham2}. In order to also avoid the eigenstates of the Hamiltonian, which give an unphysical total spin number, we only pick up the exact eigenstates with conserved charges \cite{github}.

Fig.~\ref{fig:Orb_Re} shows that the corresponding ground state energies and the first excited state energies vary with the molecular bond length. In fig.~\ref{fig:Orb_Re} (a), the ground state and the first excited state's energy with less than 5. The corresponding relative errors are shown in fig.~\ref{fig:Orb_Re} (b). As the radius increases, the relative error of the result with less than 5 molecular orbitals becomes lower. Although the result with 5 molecular orbitals is a bit off from the exact one at $r=0.4$ atomic unit, overall, the relative error of 5 molecular orbitals' result is lower than $10^{-3}$. Therefore, we ignore one molecular orbital which corresponds to core orbital of $LiH$ compound for the VQE simulation.

 \subsection{Optimizer Comparison}
Before comparing various VQE methods, we assess the effectiveness of different optimizers—GD, QNG, and Adam—in obtaining the first excited state of $LiH$ molecules. Here, we select FS-VQE to simulate the excited state of $LiH$ since it's intricate to implement Pennylane's QNG function with SSVQE or VQD. For the hyperparameters, we maintain a consistent learning rate of $0.07$ across all optimizers and fix the bond length at $l=1.6$ Å and $\omega=-7.8$ Ha in eq.~\ref{eq: fold}. Additionally, in the FS-VQE simulation, we employ three layers of spin-restricted QCCSD with initial parameters initialized to zero. Moreover, we establish a stopping criterion where iteration continues until the difference between the current iteration's cost function and the previous iteration falls below the convergence threshold of $10^{-6}$, or until the optimization process reaches the 400th iteration.

Fig.~\ref{fig:OPT} depicts the cost function value varying with the iterations for three optimizers. GD's cost function decays more and more slowly and stops at the 157th iteration. The final value obtained using GD still falls $0.02247$ short of the target value. 
With QNG, the cost function decreases drastically for the first few iterations but slows down the decay after 50 iterations. Finally, the cost function stops at the 258th iteration. The final value obtained using QNG is $0.00429$ away from the target. Although the Adam optimizer's cost function exhibits significant fluctuation during the initial iterations, it stabilizes and steadily decreases thereafter, converging by the 134th iteration. The final value is merely $2.14 \times 10^{-5}$ from the target value. The Adam optimizer's performance surpasses that of two other optimizers. Notably, while the number of circuits required for Adam equals that of GD, the QNG demands additional circuits for evaluating the quantum geometric tensor as outlined in eq~\ref{eq: GM}. Considering efficiency as a pivotal factor, Adam stands out as the most suitable optimizer for training $LiH$'s excited states. Consequently, we opt to utilize Adam as the optimizer for the remainder of our calculations. 

\subsection{Results of Comparison}

To ensure a fair comparison between VQD and SSVQE simulations, we employ a spin-restricted QCCSD ansatz and Adam optimizer for both methods. Additionally, we maintain identical learning rates of $0.3$ and initial states for each. The only distinction lies in the number of layers of the spin-restricted QCCSD. Given VQD's approach of optimizing states individually, we employ 2 layers of QCCSD for the ground state, 3 layers for the first excited state, and 4 layers for the second excited state. Conversely, for SSVQE, we utilize 4 layers of QCCSD as the common parameterized unitary for all states. Moreover, we also have the same stopping criterion with a convergence threshold of $10^{-5}$ here.

\begin{figure}[t]
     \includegraphics[width=0.5\textwidth]{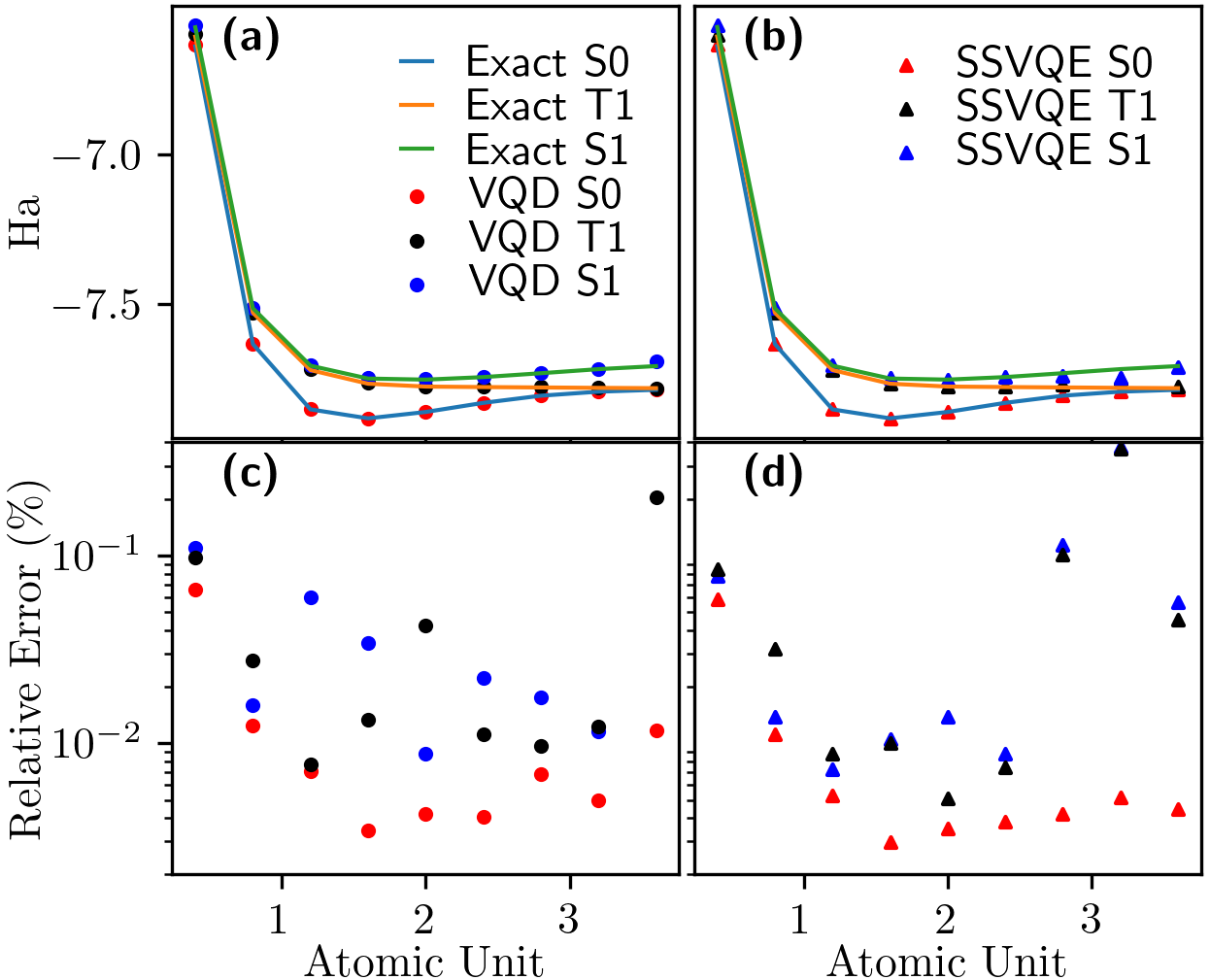}
     \centering
     \caption{The Upper panel: the LiH's energy calculation varying with bond length for singlet ground state (S0), triplet first excited states (T1), and singlet first excited state (S1) labeled as \textbf{(a)} circle dots using VQD, solid lines using exact diagonalization with 6 orbitals, and \textbf{(b)} triangle dots using SSVQE. The lower panel: the corresponding relative error labeled as \textbf{(c)} circle dots using VQD and \textbf{(d)} triangle dots using SSVQE. The relative error here is defined as $\left|\frac{\braket{H}_{\text{VQE}}-\braket{H}_{\text{Exact}}}{\braket{H}_{\text{Exact}}}\right|$.}
     \label{fig:compar_1}
 \end{figure}

 The results are distilled into figs.~\ref{fig:compar_1} visualizing the energy of the ground state and first two excited states with $S_z=0$ as $LiH$ bond length varies.
On the other hand, fig.~\ref{fig:compar_1} (a) shows that VQD energies calculation for the singlet ground state (S0), triplet first excited state (T1), and singlet first excited state (S1) match the exact numerical calculation. The SSVQE's energy calculation shown in fig~\ref{fig:compar_1} (b) also fits the exact one except for the first singlet excited state energies with bond length $l=2.8\textrm{Å}$ and $l=3.2\textrm{Å}$. Figures \ref{fig:compar_1}(c) and \ref{fig:compar_1}(d) respectively display the relative errors of the S0, S1, and T1 states with $S_z=0$ for VQD and SSVQE. In the case of VQD, the relative errors are higher at shorter bond lengths due to the exclusion of the core molecular orbital. However, overall, the relative errors of the energy states remain below $0.1 \%$, except for the excited state at $3.6\textrm{Å}$. Conversely, for SSVQE, the relative errors of the S1 and T1 states' energy, with bond lengths longer than $2.8\textrm{Å}$, range from $0.07 \%$ to $0.4 \%$, which is higher than others except for the relative errors of energies at $0.4\textrm{Å}$. Although most of VQD and SSVQE's relative errors are below $0.1\%$, as table \ref{table:data} shows, they require $448 \sim 896$ $2$-qubits gates, $800 \sim 1600$ $1$-qubit gates for the parameterized quantum circuit, which are unfriendly for NISQ devices.

To make the simulation more practical for the NISQ device, a reduction in qubits' number and number of gates is necessary. Here, we use qubit tapering to reduce the number of circuits. Using qubit tapering, the number of qubits is reduced to $6$. Moreover, to further reduce circuit depth, we also try the SE ansatz, which is easily implemented on quantum hardware, on VQD with $12$ layers for the ground state, $14$ layers for the triplet excited state, and $16$ layers for the singlet state. To get $S_z=0$ states, we set up $m_z=0$ in eq~\ref{eq:sz} for the cost function.
\begin{figure*}[htpb]
     \includegraphics[width=1\textwidth]{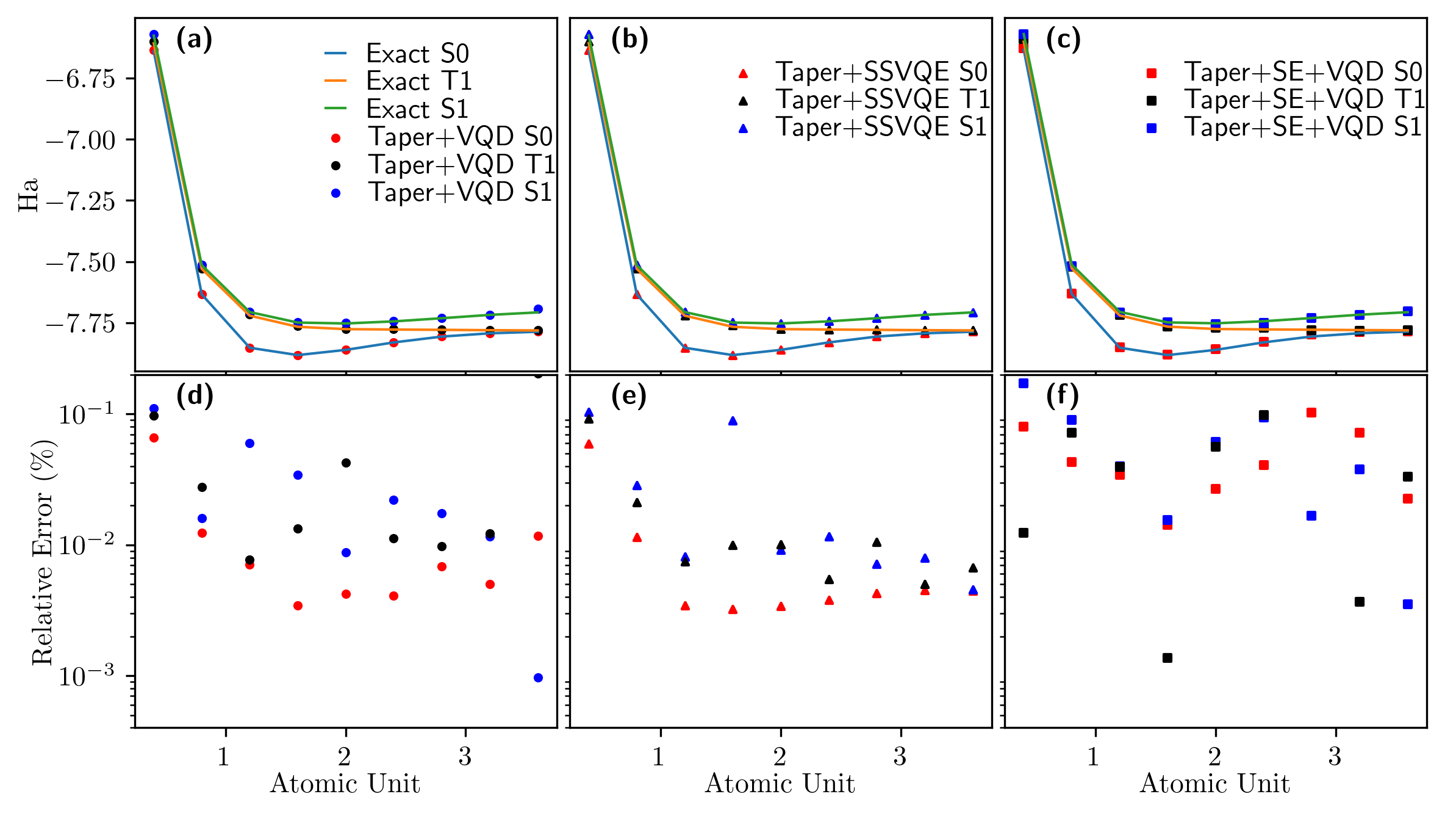}
     \centering
     \caption{The energy calculation and the corresponding relative error versus the bond length. The upper panel: S0, T1, S1's energy estimated using exact diagonalization with 6 orbitals, \textbf{(a)} VQD+Spin Restricted QCCSD+Tapering (circle dots), \textbf{(b)} SSVQE+Spin Restricted QCCSD+Tapering (triangle dots), \textbf{(c)} VQD+Strongly Entangled Ansatz+Tapering (square dots). The lower panel: the corresponding relative error of \textbf{(d)} VQD+Spin Restricted QCCSD+Tapering (circle dots), \textbf{(e)} SSVQE+Spin Restricted QCCSD+Tapering (triangle dots), \textbf{(f)} VQD+Strongly Entangled Ansatz+Tapering (square dots). }
     \label{fig:compar_2}
 \end{figure*}

 Fig.~\ref{fig:compar_2} shows the results of VQD+spin restricted QCCSD, VQD+SE ansatz, and SSVQE+spin restricted QCCSD with qubit tapering. In the upper panel of figs.~\ref{fig:compar_2}, the energies obtained from all three methods match the exact one very well. In figs.~\ref{fig:compar_2} (d), with the qubit tapering technique, the relative error remains the same in fig.~\ref{fig:compar_1} (c), but, as table \ref{table:data} shows, with qubit tapering technique, the number of $2$-qubit gates ($1$-qubit gates) is reduced to $246 \sim 492$ ($446 \sim 892$). The circuit depth is reduced by $39.85 \%$. Moreover, the number of trainable parameters is also significantly reduced by $70 \%$. This means there are much less evaluations for the loss function's gradient. Figs.~\ref{fig:compar_2} (e) shows that the qubit tapering technique improves the energy result at the bond length longer than $2.4\textrm{Å}$. Except for bond length $l=0.4\textrm{Å}$, the relative errors are less than $0.1 \%$. In figs.~\ref{fig:compar_2} (f), although the VQD+strongly entangled ansatz's relative errors are not lower than VQD+QCCSD's, in table~\ref{table:data}, the circuit depth to implement $16$ layers SE ansatz is just $144$ which is less than $429$, the 2 layers of VQD+QCCSD's circuit depth. However, the number of trainable parameters for 12 layers of SE is more than $2$ times that of 4 layers of QCCSD. That means that more evaluations are required for the gradient of the loss function. Furthermore, unlike spin restricted QCCSD, which searches for the optimal parameters within the spin sector, SE finds the optimal one within the entire Hilbert space. It turns out that, compared to spin restricted QCCSD, SE requires more iterations to get the training result to converge, as shown in table~\ref{table:data}. We also try $16$ layers of QCCSD for SSVQE, however, the results are not very accurate.

\begin{table}[t]
\begin{tabular}{p{1.2cm}|p{1.1cm}p{0.9cm}p{1cm}p{1.1cm}p{1.1cm}}
\hline
 &VQD& SSVQE& VQD+Tap&SSVQE+Tap&VQD+Tap\\ &+QCCSD &+QCCSD&+QCCSD&+QCCSD&+SE\\
\hline
\# 1q gates &S0:800&1600&S0:446&892&S0:216\\
 &T1:1200&&T1:669&&T1:252\\
 &S1:1600&&S1:892&&S1:288\\
\hline
\# 2q gates &S0:448&896&S0:246&492&S0:72\\
&T1:672&&T1:369&&T1:84\\
&S1:896&&S1:492&&S1:96\\
\hline
depth &S0:713&1425&S0:429&857&S0:108\\
&T1:1069&&T1:643&&T1:126\\
&S1:1425&&S1:857&&S1:144\\
\hline
\# iters &S0:66.44&114.11&S0:66.44&114.11&S0:290.44\\
&T1:66.56&&T1:68.56&&T1:209.56\\
&S1:71.89&&S1:71.89&&S1:163.78\\
\hline
\# params &S0:48&96&S0:20&40&S0:216\\
&T1:72&&T1:30&&T1:252\\
&S1:96&&S1:40&&S1:288\\
\hline 
\end{tabular}
\centering
\caption{The quantum parameterized circuit's detail (circuit depth, 1 qubit and 2 qubit gates' number, the number of parameters) and the averaging iteration to get the results (average over 9 bond length) for each method \cite{github}.}
\label{table:data}
\end{table}

\subsection{Explore Highly Excited States}

To investigate the highly excited states, we can use the FS-VQE or adaptive VQE-X \cite{Zhang21}. All of them cannot find more than one highly excited state at once. Here, we propose a mixture of FS-VQE and VQD or SSVQE, which can be called FS-VQD and FS-SSVQE. Once the $\omega$ is determined, FS-VQD or FS-SSVQE can find the highly excited states near $\omega$. The corresponding FS-VQD's cost function is
\begin{equation}
\mathcal{L}(\boldsymbol{\theta}_k)=\bra{\psi(\boldsymbol{\theta}_k)}\left(H-\omega\right)^{2}\ket{\psi(\boldsymbol{\theta}_k)}+\sum_{i=0}^{k-1}\beta_{i}|\braket{\psi(\boldsymbol{\theta}_k)|\psi(\boldsymbol{\theta}^*_i)}|^{2}
\label{eq: fs-vqd}
\end{equation}
where the first term is from FS-VQE and the second term comes from VQD. $\beta$ should be the value larger than the energy discrepancy. 
And we can use it as the absolute loss function by applying the square root to the first term of Eq.~\ref{eq: fs-vqd}. The loss function of FS-SSVQE is
\begin{equation}
\mathcal{L}_{w}(\boldsymbol{\theta})=\sum_{j=0}^{k}w_j\bra{\psi(\boldsymbol{\theta})}\left(H-\omega\right)^{2}\ket{\psi(\boldsymbol{\theta})},
\label{eq: fs-ss-vqe}
\end{equation}
where $w_j\in\left[0,1\right]$, and  to convert this into an absolute loss function, it is necessary to apply a square root to each expectation value
Here, we select a chemical molecule $H_4$, which requires $8$ qubits representing $8$ spin-orbitals from $H_4$, for testing FS-VQD and FS-SSVQE. Our goal is to find the three highly excited states with energy close to $-1.0$ Ha. For the hyperparameters setup for the VQE, we choose the learning rate $l_r=0.3$, $\beta=5$ for FS-VQD, and $w_j=0.4$ with all $i$ for FS-SSVQE. Since last subsection shows that the cost function with QCCSD ansatz converges faster than that with SE, we employ 5 layers of spin restricted QCCSD as the ansatz with trainable parameters setting as zero initially. We also implement the stopping criterion with a convergence threshold of $10^{-5}$ here.

\begin{figure}[t]
     \includegraphics[width=0.5\textwidth]{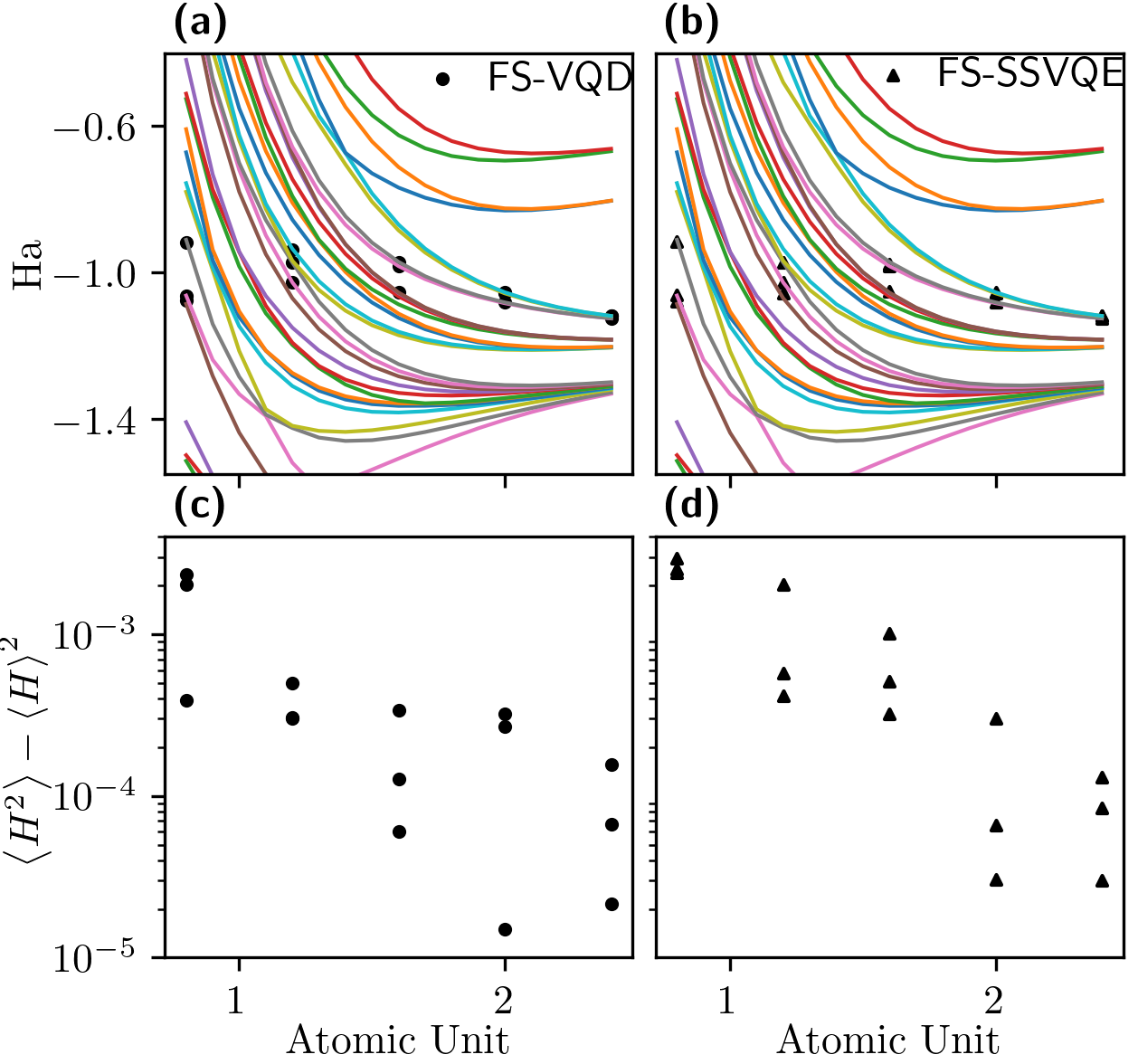}
     \centering
     \caption{$H_4$'s highly excited states' energy varying with the bond length (Upper panel) and the corresponding energy variance (Lower panel). The upper panel: The highly excited states' energies near $-1.0$ Ha calculation using: \textbf{(a)} FS-VQD labeled as circle dots and \textbf{(b)} FS-SSVQE labeled as triangle dots. The lower panel: The energy variance of the excited state obtained from \textbf{(c)} FS-VQD labeled as circle dots and \textbf{(d)} FS-SSVQE labeled as triangle dots.}
     \label{fig:HES}
 \end{figure}

Figs.~\ref{fig:HES} manifest highly excited states' results from FS-VQD and FS-SSVQE. In figs.~\ref{fig:HES} (a) (b), the highly excited states near $-1.0$ Ha are captured well, respectively, by the FS-VQD and FS-SSVQE. Because there are many excited states near $E=-1.0$ Ha, it is hard to use relative error to describe how good FS-VQD and FS-SSVQE's performances are. Instead, we use energy variance as the measure to see how precise the FS-VQD and FS-SSVQE capture the highly excited eigenstates. If the energy variance is zero, the corresponding state is exactly an eigenstate. In fig.~\ref{fig:HES} (c), the energy variance gained from FS-VQD is below $10^{-3}$ except for the two highly excited states at bond length $l=0.8\textrm{Å}$. As the bond length increases, the energy variance also decreases. On the other hand, in fig.~\ref{fig:HES} (d), the energy fluctuations gotten from FS-SSVQE are a bit higher than the FS-VQD's except for the states at bond length $l=2.0\textrm{Å}$. Compared to the low-lying excited states, the optimization for highly excited states demands much more iterations to get the loss function converged.           



\section{Conclusion\label{sec4}}

We conducted a comprehensive analysis, comparing the performance of VQD and SSVQE methods in determining the low-lying excited states of $LiH$. In terms of accuracy, the performance of VQD and SSVQE are almost the same. However, from an efficiency standpoint, SSVQE stands out by allowing for the determination of all low-lying excited states through only a single optimization process. Furthermore, our investigation evaluates the effectiveness of various optimizers, namely GD, QNG, and Adam, in obtaining LiH's first excited state. The Adam optimizer demonstrates superior efficiency, requiring the fewest iterations to achieve the desired excited state. We propose a novel method combining FS-VQE with either VQD or SSVQE to explore highly excited states. This method allows the exploration of a single excited state and uncovers multiple highly excited states near a specific energy level simultaneously. We also test their efficacy to get $H_4$'s highly excited states near $-1.0$ Ha. FS-VQD spends less iterations than FS-SSVQE to get highly excited states although FS-SSVQE can produce all of highly excited states near $\omega$ with only one optimization process. 

One of FS-VQD and FS-SSVQE's applications is to search for the many-body scar states\cite{Turner2018,Turner2018_1,Moudgalya2018} in the spectrum like paper\cite{Cenedese2024} or prepare the many-body scar states for the specific Hamiltonian using shallow circuits. While existing implementations have explored the preparation of many body scar states using quantum computers\cite{Gustafson2023}, employing shallow circuit VQE to prepare many-body scar states can help us understand the circuit depth required for preparing the scar states, especially for those of which entanglement entropy satisfying $O(\log L)$ \cite{Schecter2019}. 

Moreover, all of our comparisons can be extended to larger molecules. It is crucial to investigate the effectiveness of VQD and SSVQE methodologies when applied to larger chemical molecules, aiming to uncover more profound insights into their performance. Furthermore, optimizing these methodologies for larger molecules warrants attention, with potential scrutiny on the effectiveness of optimizers such as GD, QNG, and Adam. While prior studies have indicated QNG's superiority over Adam for systems exceeding a size of 20, these assessments have been limited to XXZ model's ground state tasks. We leave those interesting topics for future research.

\section*{Data Availability}
The data that support the findings of this study are openly available in Qhack2024-project repository at \url{https://github.com/ichen17/Qhack2024-project/}.
\section*{Acknowledgement}
The authors acknowledge valuable discussions with Ivana Kurečić, Kl\'ee Pollock, and Akhil Francis. This project won first prize in the ``Bridging the Gap'' category at QHack 2024. The authors appreciate Xanadu Quantum Technologies Inc. for holding the QHack $2024$. Work by I.-C.C. and J.S. was supported by the U.S. Department of Energy (DOE), Office of Science, Basic Energy Sciences, Materials Science and Engineering Division, including the grant of computer time at the National Energy Research Scientific Computing Center (NERSC) in Berkeley, California. 
This work was partially conducted (I.-C.C., J.S.) at Ames National Laboratory which is operated for the U.S. DOE by Iowa State University under Contract No. DE-AC02-07CH11358. J.S. was supported by the U.S. Department of Energy, Office of Science, National Quantum Information Science Research Centers, Superconducting Quantum Materials and Systems Center (SQMS) under Contract No. DE-AC02-07CH11359.


\bibliographystyle{IEEEtran}

\bibliography{refs}
\end{document}